# Estimation of urinary stone composition by automated processing of CT images.

Grégoire Chevreau [1,2], Jocelyne Troccaz [2], Pierre Conort [1], Raphaëlle Renard-Penna [3], Alain Mallet [4], Michel Daudon [5], Pierre Mozer [1].

1. Department of Urology - Pitié-Salpêtrière Hospital, Paris, France
2. GMCAO Team – TIMC/IMAG Laboratory, Joseph Fourier University - CNRS (UMR 5525), Grenoble, France.
3. Department of Radiology – Pitié-Salpêtrière Hospital, Paris, France.
4. Department of Biostatistics - Pitié-Salpêtrière Hospital, Paris, France.
5. Department of Biochemistry A – Necker Hospital, Paris, France.

***Objective***: Developing an automated tool for routine clinical practice to estimate urinary stone composition from CT images based on the density of all constituent voxels.

***Material and methods***: 118 stones for which the composition had been determined by infrared spectroscopy were placed in a helical CT scanner. A standard acquisition, low dose and high dose acquisitions were performed.
All voxels constituting each stone were automatically selected. A dissimilarity index evaluating variations of density around each voxel was created in order to minimize partial volume effects: stone composition was established on the basis of voxel density of homogeneous zones.

***Results***: Stone composition was determined in 52% of cases. Sensitivities for each compound were: uric acid: 65%, struvite: 19%, cystine: 78%, carbapatite: 33.5%, calcium oxalate dihydrate: 57%, calcium oxalate monohydrate: 66.5%, brushite: 75%.
Low-dose acquisition did not lower the performances ($p<0.05$).

***Conclusion***: This entirely automated approach eliminates manual intervention on the images by the radiologist while providing identical performances including for low-dose protocols.

# Introduction

In order to select the most appropriate treatment for each patient, the urologist tries to obtain information characterizing the stone(s) to be treated. Following demonstration of the value of CT in the assessment of urinary stone disease [1], it has become an essential examination for the management of these patients. It is therefore used routinely to precisely determine the site, dimensions and position of the stones. However, determination of their mineral composition is more difficult. Although it has been demonstrated experimentally, on a micro CT scanner [2], that X rays are able to distinguish between the various mineral compounds, in routine clinical practice stone composition can only be estimated from CT images by manual definition of one or several Regions Of Interest (ROI) followed by assessment of their voxel densities [3].

This approach provides unreliable results, as the ROI examines only a sample of a stone, which is usually heterogeneous, and because of partial volume effects between the voxels selected. Finally, this approach requires time-consuming expert human intervention associated with probably imperfect reproducibility.

In order to overcome these problems, we propose an automated image processing method based on analysis of the entire stone by identifying homogeneous zones within the stone and determining their mineral composition.

This approach was tested on *ex vivo* stones with a known composition, according to various CT acquisition parameters in order to evaluate the impact of irradiation modifications on the precision of this technique.

## 1. Materials and methods

A total of 118 stones extracted by endoscopic or percutaneous surgery for which the composition was determined by infrared spectroscopy were selected. These stones presented a range of compositions and dimensions in order to be representative of routine clinical practice.

The stones consisted of pure or mixed forms of the following biochemical compounds: uric acid (UA), calcium oxalate monohydrate (C1) or dihydrate (C2), struvite, carbapatite (CA), cystine (Cys), and brushite (Br). The stone diameter ranged from 1 mm to about 20 mm.

Each stone was placed in a Plexiglas jar measuring 3 cm in diameter and 5 cm high and included in a fat that is solid at room temperature (Végétaline®) to avoid any contact between the stone and the edge of the jar.

The 118 Plexiglas jars were then placed in a single box that was introduced into the Phillips Brilliance 64 CT scanner composed of 64 rows of detectors. Standard acquisition parameters for the detection of urinary tract stones in patients were used, i.e. 120 kV, 250 mAs, Pitch: 0.641, collimation: 64 x 0.625. The voxel size obtained was 0.75 x 1 x 1 mm.

In order to assess the performances of this approach at various tube currents, 5 successive acquisitions were performed. A standard acquisition was therefore performed at 250 mAs, a high-dose acquisition was performed at 500 mAs and three low-dose acquisitions were performed at 100, 80 and 50 mAs.

Each stone can be extracted from the CT volume by a simple image processing algorithm (Figure 1). As the inclusion medium in which the stones were placed presents a negative density of about -100 Hounsfield Units (HU) and as all mineral components of a stone have a positive density in HU, extraction of each stone was therefore performed using a simple 0 HU cut-off.

For each acquisition, all stones were therefore extracted automatically and the density (in HU) and 3D coordinates in the CT image volume were determined for each voxel (3D equivalent of a pixel). All these steps were performed with Analyse 7.0 © software [4].

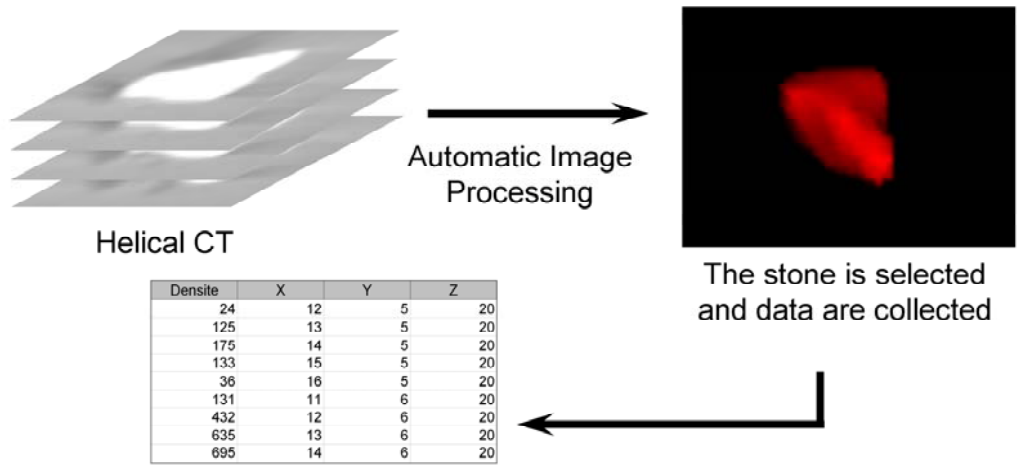

*Figure 1: CT image processing isolates the stone, allowing analysis of each voxel.*

1.1     Identification of homogeneous zones:

Due to the relatively large voxel size (0.75 x 1 x 1 mm), some voxels overlapped two different mineral compounds, inevitably resulting in partial volume effects within the stone. The density value for these voxels is therefore a weighted mean of the densities of the mineral compounds present in the voxel. These noninformative voxels therefore need to be separated from those situated in homogeneous zones of the stone and which provide more relevant information.

A density dissimilarity index between neighbouring voxels was therefore defined. Neighbouring voxels on a digital image can be defined according to various types of connectivity: 2D (4- and 8-connectivity) and 3D (6-, 18- and 26-connectivity). In 3D for example, neighbour voxels can be considered to be those which have a common face (6-connectivity), a common face or a common edge (18-connectivity), or a common face, a common edge or a common vertex (26-connectivity): Figure 2.

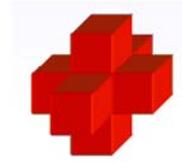

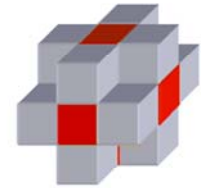

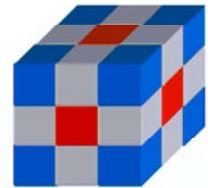

*Figure 2: The various types of 3D connectivity*

All these data (coordinates, density, dissimilarity index) and the composition of each stone were exported to a database. A total of 390,000 voxels were analysed, i.e. almost 78,000 voxels per acquisition.

The CT attenuation in HU of informative voxels (i.e. those situated within homogeneous zones) was then correlated with the known stone composition determined by infrared spectroscopy, by determining the density distribution of each mineral compound by a Bayesian approach.

As only a few of the stones had a pure mineral composition, all available stones were used to establish the densities of each mineral compound. To decrease density overlap of the various compounds, a repeated classification method designed to minimize assignment of voxels to a mineral compound not present in the stone was implemented. A density distribution scale was therefore established for each mineral compound.

To increase the robustness of the results, cross-validation was performed by using all voxels of the database apart from those of the stone for which the composition had to be estimated. Stone voxels were therefore excluded from the voxels used for the algorithm learning step.

Finally, the main composition of the stone (predominant compound in percentage of voxels) estimated by voxel analysis was compared to the composition determined by infrared spectroscopy (predominant compound as well for mixed stones).

### 1.2    Study of the effect of tube current

The McNemar Chi-square test was used to compare the performances of this tool to correctly classify a stone according to various acquisition modalities (Standard Dose, High Dose and Low Dose).

## 2. Results

2.1 Density of the various mineral compounds and stone composition

With a dissimilarity index less than 0.3, as discussed below, the densities obtained for each mineral compound by this method are represented in Table 1 and Figure 3.

| Table 1 -Density distribution of the various mineral compounds in Hounsfield Units (HU) | |
|---|---|
| Mineral compound | Density: mean ± standard deviation (HU) |
| Uric Acid | 477 +/- 108 |
| Struvite | 613 +/- 67 |
| Cystine | 713 +/- 66 |
| Carbapatite | 948 +/- 109 |
| C2 | 1139 +/- 40 |
| C1 | 1305 +/- 110 |
| Brushite | 1610 +/- 100 |
| C1 and C2: Calcium oxalate monohydrate and dihydrate. | |

*Table 1 : CT attenuation values of the various mineral compounds in Hounsfield Units (HU). C1 and C2: Calcium oxalate monohydrate and dihydrate.*

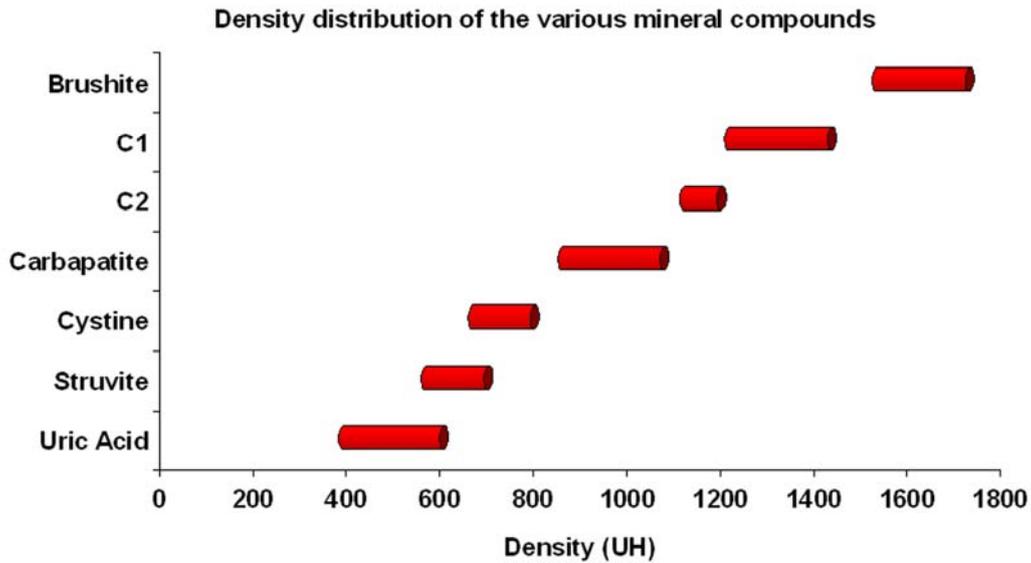

*Figure 3: CT attenuation values of the various mineral compounds*

The attenuation values characterizing each mineral compound allowed us to accurately determine the predominant compound present in each stone in 52% of cases.

Sensitivity and specificity values for each mineral compound were 65% and 92% for uric acid, 19% and 94% for Struvite, 78% and 97% for cystine, 33% and 89% for Carbapatite, 57% and 86% for C2, 67% and 89% for C1, 75 and 96% for Brushite, respectively.

These values are shown in Table 2.

Table 2: Classification of 95 stones presenting voxels with a dissimilarity index less than 0.3

| CT-estimated composition | Infrared-determined composition | | | | | | | |
|---|---|---|---|---|---|---|---|---|
| | UA | Struvite | Cys | CA | C2 | C1 | Br | Total |
| Uric Acid | 11 | 6 | | | | | | 17 |
| Struvite | 3 | 3 | 2 | | | | | 8 |
| Cys | 1 | 2 | 7 | | | | | 10 |
| CA | 2 | 4 | | 7 | 2 | | | 15 |
| C2 | | 1 | | 5 | 4 | 6 | | 16 |
| C1 | | | | 6 | 1 | 14 | 1 | 22 |
| Br | | | | 3 | | 1 | 3 | 7 |
| Total | 17 | 16 | 9 | 21 | 7 | 21 | 4 | 95 |
| Se (%) | 64.71 | 18.75 | 77.78 | 33.33 | 57.14 | 66.67 | 75.00 | |
| Sp (%) | 92.31 | 93.67 | 96.51 | 89.19 | 86.36 | 89.19 | 95.60 | |

UA : Uric Acid, Cys : Cystine, CA : Carbapatite, C1 : Calcium Oxalate Monohydrate, C2 : Calcium Oxalate Dihydrate, Br : Brushite, Se : Sensitivity, Sp : Specificity

*Table 2: Classification of 95 stones presenting voxels with a dissimilarity index less than 0.3*

2.2 Modification of tube current

No significant difference in terms of stone classification was observed between the various types of CT acquisition: low-dose acquisition provided the same performances as standard or high-dose acquisition.

The sensitivity of the algorithm to classify the main constituent of the stone was as follows: 50% at 50 mAs, 49% at 80 mAs, 51% at 100 mAs, 52% at 250 mAs, and 49% at 500 mAs (no significant difference; $p< 0.05$).

No significant difference was observed for the sensitivities or specificities for each mineral compound as a function of the various acquisitions ($p<0.05$).

## 3. Discussion

This study addressed two different issues: the possibility to determine the mineral composition of a stone on the basis of the HU density of voxels considered to be informative during standard CT acquisition and the impact of low-dose acquisition on the capacity of this tool to determine mineral composition.

3.1 Density of the various mineral compounds and stone composition

Based on analysis of our data, the dissimilarity index was defined for a cut-off of 0.3 for neighboring voxels defined by 3D 6-connectivity. This choice allowed the use of three-dimensional data and the selection of a sufficient number of voxels and stones to allow statistical analysis.

In the present series of 118 stones, 23 did not present voxels with a dissimilarity index less than 0.3 and were therefore not selected for analysis of stone composition. These 23 stones had a mean diameter of 3.1 mm (range: 1.0 to 4.4 mm). In clinical practice, the majority of stones in this size range are eliminated spontaneously. This study was not designed to determine the composition of such small stones.

The proposed tool was able to estimate the main mineral composition of the stone in 52% of cases. The composition of a stone is important for the urologist in order to select the most appropriate treatment according to the site, size and supposed fragility of the stone to extracorporeal shock-wave lithotripsy (ESWL). On the basis of previous studies [5-7], urinary stones were classified into three main groups:

- Uric acid stones which can be treated medically.
- Stones that can be easily fragmented by ESWL: Struvite, C2 and CA
- Stones resistant to ESWL, requiring endourological management: Cystine, C1 and Brushite.

By using this classification, 66% of stones are assigned to the right treatment group. Table 3 presents the results of this classification.

It is difficult to compare the results of this study with those published in the literature: a large number of published studies have described the densities of each mineral compound [3, 8-10]. As CT acquisition parameters vary from study to study, it is difficult to compare the densities recorded for each mineral compound, bearing in mind, as demonstrated by Saw et al [11], that the variation of collimation induces marked variability of density measurement.

The study reported here can be compared with that performed by Bellin et al [12] in 2004, as the stones analyzed were mostly the same in the two series. However, CT acquisition modalities and the approaches used for stone classification differed between these two studies. The performances of each approach can therefore be compared, but not the density of each mineral compound.

Bellin et al [12] classified stones according to parameters obtained by density measurements in a ROI: highest CT-attenuation value, highest CT-attenuation value/area ratio and a visual density index established by a radiologist experienced in urological disease.

Compared to the approach used by Bellin et al [12], our approach showed a similar ability to correctly classify the major mineral compound. The advantage of our method is that it does not require image analysis by an expert radiologist to establish the ROI or to evaluate the visual density index. Our approach is also perfectly reproducible and provides data on the volume of the stone, a more relevant parameter than its long axis to guide the choice of the most appropriate treatment of the stone.

Table 3: Combined classification of 95 stones presenting voxels with a dissimilarity index less than 0.3

| CT-estimated composition | Infrared-determined composition | | | total |
|---|---|---|---|---|
| | UA | Cystine, C1 and Brushite | Struvite, C2 and CA | |
| UA | 11 | | 6 | 17 |
| Cystine, C1 and Brushite | 1 | 26 | 12 | 39 |
| Struvite, CA and C2 | 5 | 8 | 26 | 39 |
| Total | 17 | 34 | 44 | 95 |
| Se (%) | 64.71 | 76.47 | 59.09 | |
| Sp (%) | 92.31 | 78.69 | 74.51 | |

UA : Uric Acid, C1 : Calcium Oxalate Monohydrate, C2 : Calcium Oxalate Dihydrate, CA : Carbapatite, Se : Sensitivity, Sp : Specificity.

*Table 3: Combined classification of 95 stones presenting voxels with a dissimilarity index less than 0.3*

3.2 Modification of tube current

Decreasing patient irradiation by decreasing the tube current did not alter the precision of this tool and an increased irradiation dose did not improve the quality.

The use of this stone classification tool therefore does not require high-dose irradiation or any complementary acquisition. The results even encourage us to reduce the irradiation delivered to patients while maintaining high spatial resolution, as also proposed by Kim et al [13] and Heneghan et al [14].

**Conclusion**

We propose an automated method to determine the composition of urinary stones based on extraction of voxels considered to be informative.

This reproducible approach that could be performed as part of routine clinical practice eliminates the need to define a ROI, which undersamples the information provided by the CT scanner and which requires analysis by a radiologist.

This method allows treatment of the stone to be adapted to its composition in 66% of cases with a moderate sensitivity but a very high specificity. Moreover, the performances of this method are not decreased when using low-dose acquisition.

This study was performed *in vitro* and an *in vivo* study is necessary to validate the method under real diagnostic conditions.

Note: This study was performed with an AFU-Pierre Fabre grant.